\documentclass{article}

\usepackage{arxiv}

\usepackage{bm,cite,color,graphicx,epstopdf,authblk,pdfpages,afterpage}
\usepackage[colorlinks=true, allcolors=blue]{hyperref}
\usepackage[table]{xcolor}
\usepackage{amsmath,mathtools,multirow,soul,subcaption}
\DeclarePairedDelimiter\ceil{\lceil}{\rceil}
\DeclarePairedDelimiter\floor{\lfloor}{\rfloor}

\title{Towards Optimal Robustness of Network Controllability:\\An Empirical Necessary Condition\thanks{This article has been published in IEEE TCAS-I: {\color{blue}Y. Lou, L. Wang, K.F. Tsang, and G. Chen, ``Towards Optimal Robustness of Network Controllability: An Empirical Necessary Condition,'' \textit{IEEE Transactions on Circuits and Systems I: Regular Papers}, 67(9): 3163–3174; \href{https://ieeexplore.ieee.org/document/9078045}{doi:10.1109/TCSI.2020.2986215} (2020)~~~ [\url{https://ieeexplore.ieee.org/document/9078045}]}}%
\thanks{Emails: \url{felix.lou@my.cityu.edu.hk}; \url{wanglin@sjtu.edu.cn}; \url{ee330015@cityu.edu.hk}; \url{eegchen@cityu.edu.hk} \textit{(Corresponding Author: Guanrong Chen)}}%
\thanks{Source code of this work is available in: \url{https://fylou.github.io/sourcecode.html}}}
\renewcommand\footnotemark{}

\author[1]{Yang~Lou}
\author[2,3]{Lin~Wang}
\author[1]{Kim~Fung~Tsang}
\author[1]{Guanrong~Chen}
\affil[1]{Department of Electrical Engineering, City University of Hong Kong, Hong Kong, China}
\affil[2]{Department of Automation, Shanghai Jiao Tong University, Shanghai 200240, China}
\affil[3]{Key Laboratory of System Control and Information Processing, Ministry of Education, Shanghai 200240, China}

\begin{document}
\maketitle
\begin{abstract}
To better understand the correlation between network topological features and the robustness of network controllability in a general setting, this paper suggests a practical approach to searching for optimal network topologies with given numbers of nodes and edges. Since theoretical analysis seems impossible at least in the present time, exhaustive search based on optimization techniques is employed, firstly for a group of small-sized networks that are realistically workable, where \textit{exhaustive} means 1) all possible network structures with the given numbers of nodes and edges are computed and compared, and 2) all possible node-removal sequences are considered. A main contribution of this paper is the observation of an empirical necessary condition (ENC) from the results of exhaustive search, which shrinks the search space to quickly find an optimal solution. ENC shows that the maximum and minimum in- and out-degrees of an optimal network structure should be almost identical, or within a very narrow range, i.e., the network should be extremely homogeneous. Edge rectification towards the satisfaction of the ENC is then designed and evaluated. Simulation results on large-sized synthetic and real-world networks verify the effectiveness of both the observed ENC and the edge rectification scheme. As more operations of edge rectification are performed, the network is getting closer to exactly satisfying the ENC, and consequently the robustness of the network controllability is enhanced towards optimum.
\end{abstract}

\keywords{Network controllability \and Robustness \and Empirical necessary condition \and Node degree \and Optimization}

\section{Introduction}
\label{sec:intro}

Complex networks have gained growing popularity and accelerating momentum since late 1990s, becoming a self-contained discipline integrating network science, systems engineering, statistical physics, applied mathematics, social sciences and the like \cite{Barabasi2016NS,Newman2010N,Chen2014Book,Chen2019Book}. The ultimate goal of understanding complex networks is to control them for utilization. In this regard, whether or not they can be controlled is essential, which leads to the fundamental concept of network controllability. Consequently, network controllability has become a focal issue in the studies of complex networks \cite{Liu2011N,Yuan2013NC,Posfai2013SR,Menichetti2014PRL,Motter15CHAOS,Wang2016AUTO,Liu2016RMP,Wang2017RSPTA,Wang2017SR,Zhang2017TAC,Xiang2019CSM}, where the concept of \textit{controllability} refers to the ability of a network being steered by external inputs from any of its initial state to any desired target state under an admissible control input within a finite duration of time.

It was shown \cite{Liu2011N} that identifying the minimum number of external control inputs (recalled driver nodes) to achieve a full control of a directed network requires searching for a maximum matching of the network, which quantifies the network \textit{structural controllability}. Practically, however, finding a maximum matching of a large-scale network is computationally time-consuming or even impossible. Along the same line of research, in \cite{Yuan2013NC}, an efficient measure to assess the \textit{state controllability} of a large-scale sparse network is suggested, based on the rank of the controllability matrix of the network.

It has been quite a long time for people to understand the intrinsic relation between topology and controllability of a general directed network. In \cite{Posfai2013SR}, it demonstrates that clustering and modularity have no discernible impact on the network controllability, while underlying degree correlations have certain effects. In \cite{Menichetti2014PRL}, it reveals that random networks of any topology are controllable by an infinitesimal fraction of driver nodes, if both of its minimum in- and out-degrees are greater than two. The underlying hierarchical structure of such a network leads to an effective random upstream (or downstream) attack, which removes the hierarchical upstream (or downstream) node of a randomly-picked one, since this attack strategy would remove more hubs than a random attack strategy does \cite{Liu2012PO}. The control centrality is defined to measure the importance of nodes \cite{Liu2012PO,Li2019ACS}, discovering that the upstream (or downstream) neighbors of a node are usually more (or less) important than the node itself. Interestingly, it is recently found that the existence of special motifs such as loops and chains is beneficial for enhancing the robustness of the network controllability \cite{Lou2018TCASI,Chen2019TCASII,Lou2019R}. The network controllability of some canonical graph models is studied and compared quite thoroughly in \cite{Wu2018JNS}. As for growing networks, the evolution of network controllability is investigated in \cite{Zhang2019PA}. Moreover, the controllability of multi-input/multi-output networked systems is studied in \cite{Wang2016AUTO,Hao2018IJRNC}, with necessary and sufficient conditions derived. A comprehensive overview of the subject is presented in a recent survey \cite{Xiang2019CSM}.

On the other hand, random failures and malicious attacks on complex networks have become concerned issues today \cite{Holme2002PRE,Shargel2003PRL,Schneider2011PNAS,Liu2012PO,Bashan2013NP,Xiao2014CPB}. To resist attacks, strong robustness is desirable and even necessary for a practical network. A measure for the network controllability is quantified by the number of external control inputs needed to recover or to retain the network controllability after the occurrence of an attack, while its robustness is quantified by a sequence of values that record the remaining levels of the network controllability after a sequence of attacks. To optimize the network robustness, one aims to enhance and maintain a highest possible \textit{connectedness} of the network against various attacks \cite{Schneider2011PNAS}. Given the degree-preserving requirement or constraint (i.e., the degree of each node remains unchanged through the process of optimization), an edge-rewiring method is proposed in \cite{Liang2015CPL} to increase the number of edges between high-degree nodes, thus generating a new network with a highest $k$-shell component. In \cite{Chan2016DMKD}, the structure of a network is modified by degree-preserving edge-rewiring, where spectral measures are used as the objective for optimization. By optimizing a specified spectral measure of the network through random edge-rewiring, the robustness of the resultant network is accordingly enhanced. It is however noticed that, although widely used as an estimator of the robustness for real-world networks, the correlation between spectral measures and the robustness remains unclear \cite{Yamashita2019COMPSAC}. Nevertheless, given a reliable predictive measure or indicator of the network robustness, optimization algorithms can be applied \cite{Liu2019ECCN,Wang2019IS}; while if there are more than one predictive measures, multi-objective optimization schemes can be applied instead \cite{Gunasekara2018MOO}. In \cite{Zeng2012PRE}, it is shown that both edge-robustness and node-robustness (i.e., the robustness against edge- and node-removals, respectively) can be enhanced simultaneously. A common observation is that heterogeneous networks with \textit{onion-like} structure are robust against attacks \cite{Schneider2011PNAS,Wu2011PRE,Tanizawa2012PRE,Hayashi2018SR}. The evolution of alternative attack and defense is studied in \cite{Ma2016PA}, where attack means edge-removal and defense means edge-replenishment. The connectedness of the largest-sized connected cluster is the typically-used measure for such robustness \cite{Schneider2011PNAS}. It should be noted that, for the study to be presented in this article, although the robustness of network connectedness has a certain positive correlation with the robustness of network controllability, they have very different measures and objectives.

Regarding the controllability of a complex network, it refers to a static property that reflects how well the network can be controlled. Yet, the robustness of network controllability is a dynamic process that reflects how well the network can maintain its controllability against destructive attacks by means of node-removal or edge-removal. Reportedly, intentional degree-based node-removal attacks in the sense of removing nodes with highest degrees are more effective than random attacks on network structural controllability over directed random-graph networks and also directed scale-free networks \cite{Pu2012PA}. In \cite{Wang2013EPL}, the optimization of robustness of controllability is transformed into the transitivity maximization for control routes. In \cite{Liang2016EPJB}, edge directionality is considered as the only operation to enhance the robustness of controllability, preserving the underlying topology meanwhile. In \cite{Chen2017PA}, the change of controllability of random networks and scale-free networks in the processes of cascading failures is studied. For networks with different topologies, the results show that the robustness of network controllability will become stronger through allocating different control inputs and edge capacities.

Both random and intentional edge-removal attacks have also been studied by many. In \cite{Nie2014PO}, for example, it shows that the intentional edge-removal attack by removing highly-loaded edges is very effective in reducing the network controllability. It is further observed, in e.g. \cite{Buldyrev2010NAT}, that intentional edge-based attacks are usually able to trigger cascading failures in scale-free networks but not necessarily in random-graph networks. These observations have motivated some recent in-depth analysis of the robustness of the network controllability \cite{Lou2018TCASI}. In this regard, both random and intentional attacks as well as both node-removal and edge-removal attacks were investigated. Specifically, for a random upstream (or downstream) attack that removes the upstream (or downstream) node of a randomly-picked one, the upstream and downstream relationship is determined by the underlying hierarchical structure of the network. This type of random attack has a non-uniform distribution, since the hub nodes are more likely being attacked in this scenario \cite{Liu2012PO}. In particular, it was found that redundant edges, which are not included in any of the maximum matchings, can be rewired or re-directed so as to possibly enlarge a maximum matching such that the needed number of driver nodes is reduced \cite{Xu2014CCDC,Hou2013ISDEA}.

Although the relations between network topology and network controllability have been investigated in some studies, there is no prominent theoretical indicator or performance index that can well describe the robustness of network controllability with a measures based on such relations. Under different attacks, the robustness of network controllability behaves differently. The nature of the attack methods leads to different measures of the \textit{importance} of a node (or an edge) in a network. Generally, degree and betweenness are two commonly-used measures for the importance \cite{Pu2012PA}.

This paper continues the above investigations to further explore the network topological properties that affect or even determine the optimal robustness of both state and structural controllability, against random node-removal attack. In this paper, with given numbers of nodes and edges, a network instance that has the optimal controllability robustness against node-removal is called an \textit{optimal instance}. A novel method is developed in the paper as follows. First, an exhaustive search is performed on a group of small-sized networks. Then, an empirical necessary condition (ENC) is observed and summarized. A simple yet effective edge rectification strategy, namely the random edge rectification (RER), is proposed for modifying the network topology to satisfy the ENC, so that the robustness of network controllability is enhanced. Similarly to \cite{Shi2013CSM}, where the optimal network topology with best possible synchronizability is observed and summarized through extensive empirical experiments, the observed ENC is confirmed by extensive numerical simulations here, since it is impossible to theoretically prove, and probably no one could do so at this time. Finally, both ENC and RER are verified by optimizations of a number of synthetic and real-world networks with different properties.

The rest of the paper is organized as follows. Section \ref{sec:robust} reviews the network controllability and its robustness against various destructive attacks. Section \ref{sec:nec} introduces the ENC and the RER. Section \ref{sec:exp} investigates both ENC and RER by extensive numerical simulations, on both synthetic and real-world networks. Section \ref{sec:end} concludes the investigation.

\section{Network Controllability and its Robustness}
\label{sec:robust}

\subsection{Network Controllability and Criteria}
\label{sub:nc}

Consider a linear time-invariant (LTI) networked system described by $ \dot{{\bf x}}=A{\bf x}+B{\bf u}$, where $A$ and $B$ are constant matrices of compatible dimensions, $\bf x$ is the state vector, and $\bf u$ is the control input. The system is \textit{state controllable\/} if and only if the controllability matrix $[B\ AB\ A^2B\ \cdots A^{N-1}B]$ has a full row-rank, where $N$ is the dimension of $A$. The concept of \textit{structural controllability\/} is a slight generalization dealing with two parameterized matrices $A$ and $B$, in which the parameters characterize the structure of the underlying networked system: if there are specific parameter values that can ensure the system described by the two parameterized matrices be state controllable, then the system is structurally controllable. In case the system is controllable, its state vector $\bf x$ can be driven from any initial state to any target state in the state space within finite time by a suitable control input $\bf u$.

The controllability of a network, or networked system, is measured by the density of the controlled nodes, $n_D$, defined by
\begin{equation}\label{eq:nd}
n_D\equiv \frac{N_D}{N}\,,
\end{equation}
where $N_D$ is the number of external control inputs (driver nodes) needed to retain the network controllability and $N$ is the network size at the current step of the process. This measure $n_D$ allows networks with different sizes can be compared. It is clear that the smaller the $n_D$ value is, the better the network controllability will be.

Generally, there are two ways to calculate the number $N_D$ of driver nodes, for structural controllability and exact (state) controllability, respectively. To introduce these two criteria, first recall from graph theory that in a directed network, a matching is a set of edges that do not share common start nodes or common end nodes. A maximum matching is a matching that contains the largest possible number of edges, which cannot be further extended in the network. A node is matched if it is the end of an edge in the matching; otherwise, it is unmatched. A perfect matching is a (maximum) matching that matches all nodes in the network.

According to the \textit{minimum inputs theorem} \cite{Liu2011N}, when a maximum matching is found, the number $N_D$ of driver nodes is determined by the number of unmatched nodes, i.e.,
\begin{equation}\label{eq:sc}
N_D=\text{max}\{1, N-|E^*|\},
\end{equation}
where $|E^*|$ is the number of nodes in the maximum matching $E^*$. If a network has a perfect matching, then the number of driver nodes is $N_D=1$ and the control input can be put at any node; otherwise, $N_D=N-|E^*|$ control inputs are needed, which should be put at those unmatched nodes.

As for exact controllability, if a network is sparse, its number $N_D$ of driver nodes can be calculated by $N_D=\text{max}\{1, N-\text{rank}(A)\}$. Here, a network is considered to be sparse if the number of edges $M$ (i.e., the number of nonzero elements of the adjacency matrix) is much smaller than the possible maximum number of edges, $M_{max}=N\cdot (N-1)$. Usually, if $M/M_{max}\leq 0.05$, then it is considered as a sparse network.

\subsection{Robustness of Network Controllability}
\label{sub:rnc}

In this paper, since only node-removal attacks is considered, the robustness of network controllability is evaluated after a node is removed, one by one, yielding a sequence of values that reflect how robust (or vulnerable) a networked system is against a destructive attack. Different attack strategies result in different damages to the network topology and also its controllability. An attack strategy is chosen according to the ``importance'' of nodes or edges in the network, where the concept of importance depends on the application in consideration. Nevertheless, no matter under which attack strategy, the measure of controllability robustness represented by the \textit{controllability curve} can be calculated as follows,
\begin{equation}\label{eq:ndi}
n_D(i)\equiv \frac{N_D(i)}{N-i}\,,\ \ i=1,2,\ldots,N-1,
\end{equation}
where $N_D(i)$ is the number of driver nodes needed to retain the network controllability after $i$ nodes have been removed, and $N$ is the original network size.

Generally, there are two types of attacks, namely intentional and random node-removal attacks. An intentional attack aims at removing a node that is the most important to maintain the network controllability; for example, the node with the largest degree or betweenness. A random attack removes a randomly-picked node at each time step. In this paper, only \textit{random node-removal\/} attacks are considered, while intentional node-removal can be similarly discussed.

A comparison of controllability robustness among different networks can be performed by either observing the controllability curves if they are distinguishable, or by comparing the robustness measure $R_c$ \cite{Schneider2011PNAS,Ruths2013CNIV}, defined as follows:
\begin{equation}\label{eq:rc}
R_c= \frac{1}{N-1} \sum_{i=1}^{N-1}n_D(i),
\end{equation}
which represents the average density of required driver nodes, throughout the entire attack process.

Note that, given a network size $N$, one can remove at most $N-1$ nodes, since the trivial case with zero node will lead to a zero-denominator in Eq. (\ref{eq:ndi}).

An \textit{optimal instance} $I^{*}$ is a network instance that requires the lowest average density of required driver nodes through the entire attack process, defined as
\begin{equation}\label{eq:oi}
I^{*}= \operatorname*{argmin}_{I\in\Omega}{R_c}\,,
\end{equation}
where $\Omega$ is the set that includes all the network instances specified by the numbers $N$ and $M$.

\section{Empirical Necessary Condition for Optimal Topology}
\label{sec:nec}

In this section, the relation between topological structure and robustness of controllability is first studied, by observing the attack simulations on some very small-sized networks. In this case, an exhaustive attack strategy can be applied. Then, the observations are summarized as the ENC, as presented by Eq. (\ref{eq:ki}) in Section \ref{sub:enc}. To rectify an arbitrarily given network towards the satisfaction of ENC, a simple rectification strategy is proposed in Section \ref{sub:erec}.

\subsection{Exhaustive Attack}
\label{sub:exat}

To understand a full picture of the controllability change under \textit{all} possible node-removal attacks, an exhaustive attack to a set of small-sized networks is first simulated and evaluated.

Given a network of $N$ nodes, there are $N!$ permutations of the node-removal sequence. Each node-removal sequence consists of $N-1$ nodes, thus the total number of node-removal sequences is $\binom{N}{N-1}\cdot (N-1)!=N!$. Note that an intentional attack (e.g., a degree-based attack) is a specific case in such (permuted) sequences. By performing the $k$-th node-removal sequence, it generates a resultant controllability curve corresponding to $n_D^{k}=[n_D^{k}(1),n_D^{k}(2),\cdots,n_D^{k}(N-1)]$. Its controllability robustness is measured by
\begin{equation}\label{eq:rci}
R_{c}^{k}= \frac{1}{N-1} \sum_{i=1}^{N-1}n_D^{k}(i).
\end{equation}

Then, its overall controllability robustness under exhaustive attack is obtained by averaging all these $N!$ controllability measures as follows,
\begin{equation}\label{eq:exh}
\langle R_c\rangle= \frac{1}{N!} \sum_{k=1}^{N!}R_{c}^{k}.
\end{equation}

The exhaustive attack strategy considers all the sequences equally. Thus, the mean resultant curve of the exhaustive attack is equivalent to the mean of random attacks when the number of repeated runs is large enough. Each node is considered of equal importance in the network controllability study, where each node $i$ ($i=1,2,\ldots,N$) has an equal probability to be removed at the $j$th ($j=1,2,\ldots,N-1$) step within any attack sequence.

\begin{figure}[htbp]
	\centering
	\includegraphics[width=.3\linewidth]{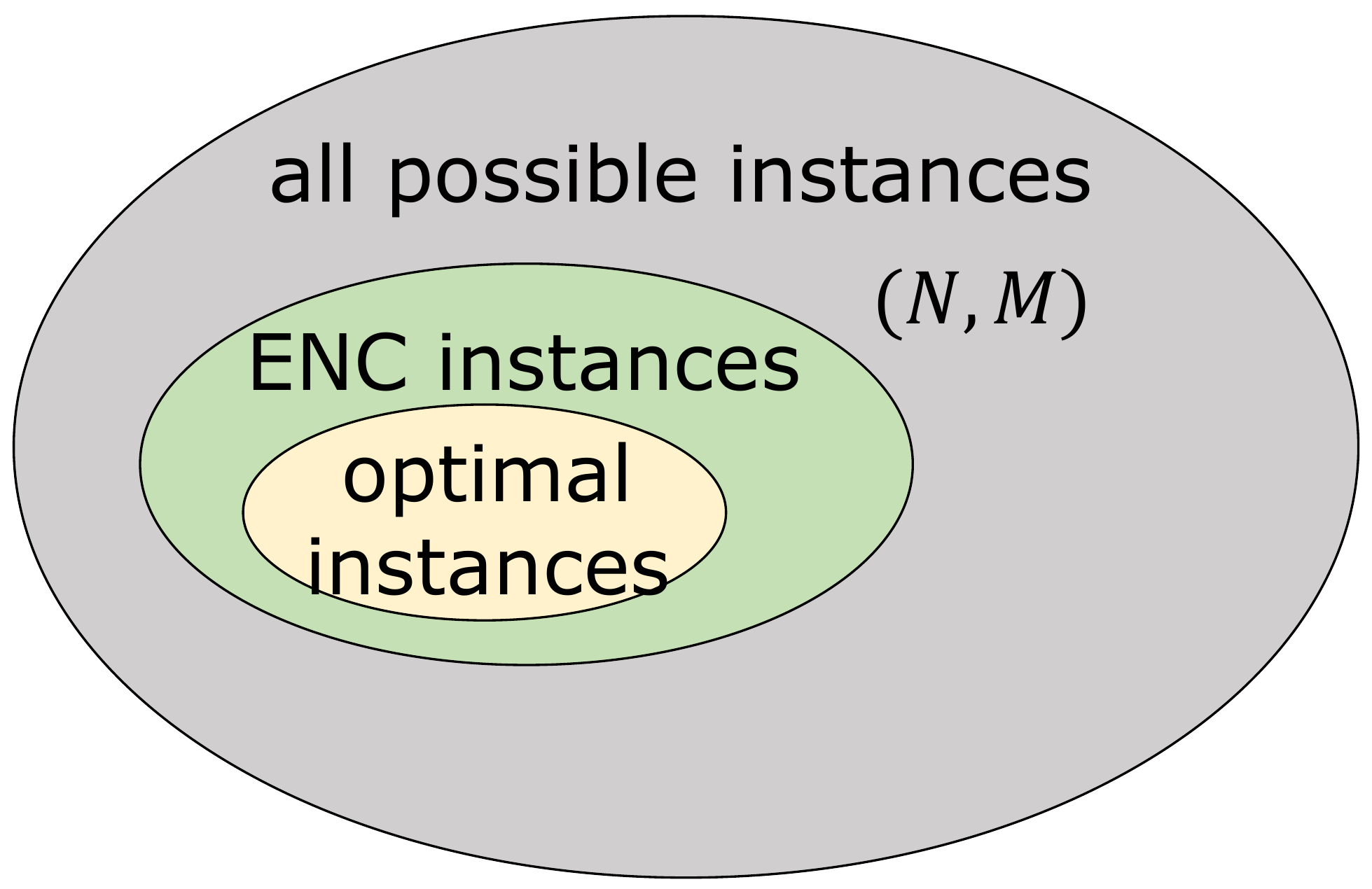}
	\caption{[Color online] Given a network with $N$ nodes and $M$ edges, the relationships among 1) all the possible instances, 2) the instances that satisfy ENC, and 3) the optimal instances.}
	\label{fig:circles}
\end{figure}

Table \ref{tab:exat} shows the results of performing the exhaustive attack on small-sized networks, where the network size is set to $4$, $5$, and $6$ only. In the table,  $\Omega$ represents the set of all possible instances with given $N$ and $M$, after filtering out isomorphs. For example, as shown in Table \ref{tab:exat} with $N=5$ and $M=10$, there are $1665$ possible combinations to form a unique network instance. For each instance, all $N!$ attack sequences are implemented and recorded. In the table, `ENC' represents the number of network instances that exactly satisfy the ENC (\ref{eq:ki}) in Section \ref{sub:enc}); $|I^*|$ represents the number of optimal instances. The topology of the optimal instance presented in Table \ref{tab:exat} can be found in Figs. S1--S3 of the Supplementary Information (SI). A phenomenon can be clearly observed from Table \ref{tab:exat} and the optimal instance presented in SI, as summarized in Fig. \ref{fig:circles}. Empirically, it is observed that the optimal instance set is a subset of the instances that exactly satisfy the ENC, which is a subset of the full set of all possible network instances with the given values of $N$ and $M$.

\begin{table}[htbp]
	\centering
	\caption{Results of performing exhaustive attack on small-sized networks. $|\Omega|$ represents the number of possible network instances with given $N$ and $M$. ENC represents the number of network instances that satisfy Eq. (\ref{eq:ki}); $|I^{*}|$ represents the number of optimal instances. The relationships of $|\Omega|$, $|I^{*}|$, and ENC sets are shown in Fig. \ref{fig:circles}.}
	\begin{tabular}{|c|c|c|c|c|c|c|c|c|c|}
		\hline
		N & M & $|\Omega|$ & ENC & $|I^{*}|$ & N & M & $|\Omega|$ & ENC & $|I^{*}|$ \\ \hline
		\multirow{8}{*}{4} & 4  & 22 & 1 & 1 & \multirow{15}{*}{5} & 5 & 108  & 1 & 1 \\ \cline{2-5} \cline{7-10}
		& 5  & 37 & 5 & 1 & & 6  & 326  & 10  & 1 \\ \cline{2-5} \cline{7-10}
		& 6  & 47 & 11& 2 & & 7  & 667  & 47  & 2 \\ \cline{2-5} \cline{7-10}
		& 7  & 38 & 5 & 1 & & 8  & 1127 & 69  & 2 \\ \cline{2-5} \cline{7-10}
		& 8  & 27 & 2 & 1 & & 9  & 1477 & 26  & 1 \\ \cline{2-5} \cline{7-10}
		& 9  & 13 & 3 & 2 & & 10 & 1665 & 5   & 1 \\ \cline{2-5} \cline{7-10}
		& 10 & 5  & 3 & 2 & & 11 & 1489 & 26  & 1 \\ \cline{2-5} \cline{7-10}
		& 11 & 1  & 1 & 1 & & 12 & 1154 & 70  & 2 \\ \cline{1-5} \cline{7-10}
		\multirow{7}{*}{6} & 6 & 582 & 1 & 1 & & 13 & 707 & 48& 2 \\ \cline{2-5} \cline{7-10}
		& 24 & 1043& 4 & 2 & & 14 & 379& 12& 1 \\ \cline{2-5} \cline{7-10}
		& 25 & 288 & 7 & 4 & & 15 & 154& 2 & 1 \\ \cline{2-5} \cline{7-10}
		& 26 & 76  & 8 & 5 & & 16 & 61 & 5 & 3 \\ \cline{2-5} \cline{7-10}
		& 27 & 17  & 5 & 4 & & 17 & 16 & 4 & 3 \\ \cline{2-5} \cline{7-10}
		& 28 & 5   & 3 & 2 & & 18 & 5  & 3 & 2 \\ \cline{2-5} \cline{7-10}
		& 29 & 1   & 1 & 1 & & 19 & 1  & 1 & 1 \\ \hline
	\end{tabular}\label{tab:exat}
\end{table}

\subsection{Empirical Necessary Condition}
\label{sub:enc}

\begin{figure}[htbp]
	\centering
	\includegraphics[width=.5\linewidth]{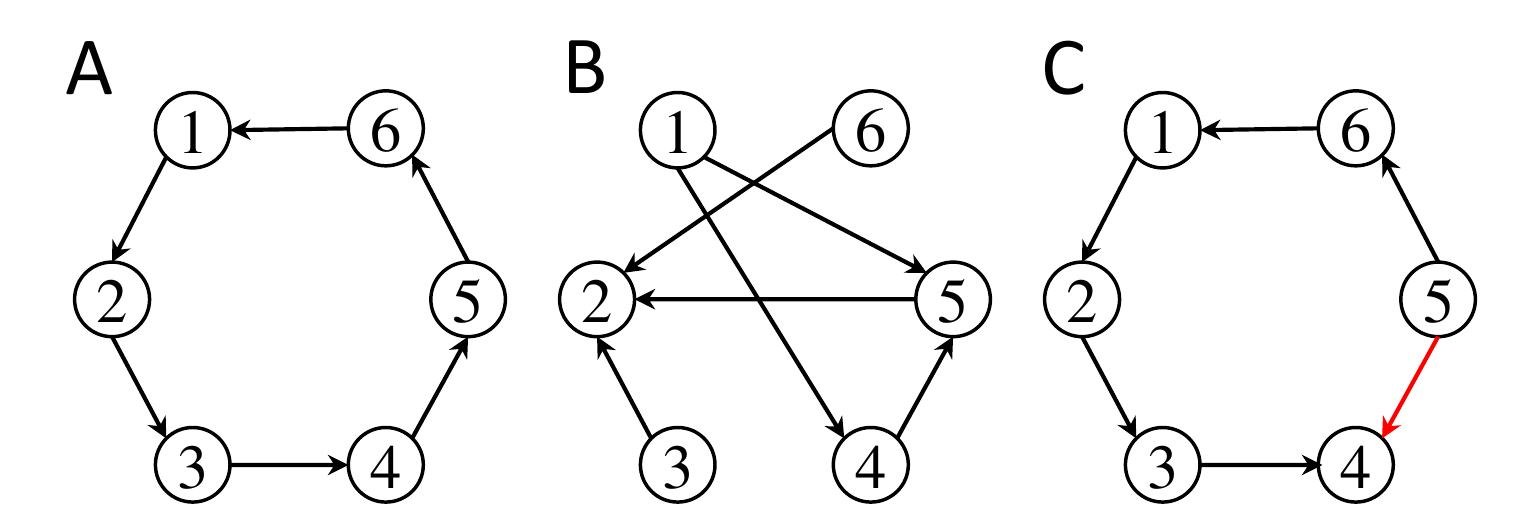}
	\caption{[Color online] Given equal numbers of nodes and edges, possible topologies include: (A) a directed global loop, (B) a tree, and (C) a ring-shaped network with a reverse edge. }
	\label{fig:eg3}
\end{figure}

It returns $N!$ controllability curves after the exhaustive attack is performed. The mean $R_c$ is calculated to be the average robustness performance. An illustrative example is shown in Fig. \ref{fig:eg3}, where there are three $6$-node networks, each has $M=6$ edges. For any $6$-node network, there are $N!=6!=720$ node-removal sequences. As shown in Table \ref{tab:eg3}, there is an $R_c$ value for each node-removal sequence. Here, the averaged value $\langle R_c\rangle$ is the overall controllability robustness measure for each network. According to Table \ref{tab:eg3}, the network shown in Fig. \ref{fig:eg3}(A) has the best controllability robustness, followed by that in Fig. \ref{fig:eg3}(C), while that in Fig. \ref{fig:eg3}(B) appears to have the worst controllability robustness among the three networks.

\begin{table}[htbp]
	\centering
	\caption{Examples with optimal instance $R_c$ values node-removal sequences (\textit{nrs})}
	\begin{tabular}{|c|c|c|c|c|}
		\hline
		\textit{nrs}& (1,2,3,4,5) & (3,2,5,6,1) & $\ldots$ & $\langle R_c\rangle$ \\ \hline
		A (loop) & 0.5833 & 0.6389 & $\ldots$ & 0.6389 \\ \hline
		B (tree)& 0.8056 & 0.7222 & $\ldots$ & 0.7667 \\ \hline
		C (ring-shaped) & 0.6667 & 0.6944 & $\ldots$ & 0.6694 \\ \hline
	\end{tabular} \label{tab:eg3}
\end{table}


Since both the number of possible network instances and the number of attack sequences increase drastically as network size increases, only very small-sized networks are examined, with results presented in Table \ref{tab:exat}. For given $N$ and $M$, the number of optimal instances is much fewer than the number of all possible instances. It can be concluded that an optimal instance has the following characteristics: 1) it contains a directed global loop that connects all the $N$ nodes; 2) both the in- and out-degrees are extremely-homogeneously distributed with extremely small differences, if any, which is consistent with the observation reported in the literature \cite{Lou2018TCASI,Lou2019R,Chen2019TCASII,Shi2013CSM} that attacks would not destroy any particularly import part of the structure due to heterogeneity. Here, the first observation may be integrated into the second. Since the in- and out-degrees of nodes in a directed loop are both extremely-homogeneously distributed, each node has in-degree one and out-degree one. This observation is also consistence with the observations reported earlier in \cite{Lou2018TCASI,Lou2019R,Chen2019TCASII}, where it was found that multiple-loop and multiple-chain structures enhance the robustness of network controllability. Therefore, based on all these observations, for a directed network with $N$ nodes and $M$ edges, the network topology with optimal robustness of controllability (against exhaustive or random node attacks) should satisfy the following condition:
\begin{equation}
\label{eq:ki}
\floor{M/N}\leq k_{i}^{in,out}\leq\ceil{M/N}\,,\ \ i=1,2,\ldots,N,
\end{equation}
where $k_{i}^{in,out}$ means both in- and out-degrees, in which as a standard notation the floor function $\floor{x}$ returns the greatest integer less than or equal to $x$, and the ceiling function $\ceil{x}$ returns the least integer greater than or equal to $x$.

Equation (\ref{eq:ki}) can be understood from two aspects: 1) $\floor{M/N}\leq k_{i}^{in,out}$ means that, for each node in the network, both its in- and out-degrees should be greater than or equal to $\floor{M/N}$, which is the integer part of the average degree $M/N$; 2) $k_{i}^{in,out}\leq\ceil{M/N}$ means its in- and out-degrees should be less than or equal to $\ceil{M/N}$. Note that, if $M$ is divisible by $N$, then $\floor{M/N}=\ceil{M/N}=M/N$. In this case, $k_{i}^{in,out}=M/N$, meaning that each node should have identical in- and out-degrees $M/N$. Otherwise, if $M$ is indivisible by $N$, then $\floor{M/N}+1=\ceil{M/N}$. In this case, the minimum of both in- and out-degrees for each node should be greater than or equal to $\floor{M/N}$, but less than or equal to $\floor{M/N}+1$.

Equation (\ref{eq:ki}) suggests that the degree distribution of the optimal instances should be allocated in a very narrow slot of the plot, i.e., it should be exactly $M/N$ if $M$ is divisible by $N$, or it should be allocated in $[\floor{M/N},\floor{M/N}+1]$ if $M$ is indivisible by $N$. For example, as illustrated by Fig. \ref{fig:eg3}, given $N$ nodes and $N$ edges, the only instance satisfying Eq. (\ref{eq:ki}) is a directed global loop (Fig. \ref{fig:eg3}(A)), where each node has in-degree one and out-degree one, and any tree structure (Fig. \ref{fig:eg3}(B)), or one with reverse edges in a ring-shaped structure (Fig. \ref{fig:eg3}(C)), does not belong to the global loop hence will significantly alter the extremely-homogeneous distribution of node degrees. This is also obvious from Table \ref{tab:exat}, where for each case of $M=N=4,5,6$, there is only one instance satisfying the ENC, which is also the optimal instance.

Empirically, it is observed that all optimal instances satisfy the ENC. But, for a network instance satisfying ENC, it is not necessarily optimal, as illustrated by Fig. \ref{fig:circles}. Therefore, this is only a \textit{necessary\/} condition.

With the restriction of ENC, the number of candidate instances in searching for optimal instances is significantly reduced. For example, as can be seen from Table \ref{tab:exat}, given $N=5$ and $M=10$, the probability for a random network instance has the optimal robustness is $1/1665$. By searching only the instances that satisfy ENC, the probability increases to $1/5$. Thus, the probability of success is largely improved and the computational cost is significantly reduced.

It can also be observed from Table \ref{tab:exat} that, as $N$ and $M$ increase, the number of instances satisfying ENC remains relatively small, compared to the total number of possible instances. Thus, with the objective of searching for the optimal instances, many instances that do not satisfy ENC can be eliminated from the candidate pool. When $N$ and $M$ are not very small, the number of possible instances could be tremendously huge. Therefore, the ENC provides an efficient means of improving the performance of searching for optimal instances from a large pool of candidates.

\subsection{Edge Rectification}
\label{sub:erec}

Since it is computational impossible to review all the possible instances when $N$ and $M$ are large, a simple yet effective strategy  called the random edge rectification (RER) is proposed here to rectify a synthetic or real-world network for its satisfaction of the ENC. It is a variant network, one with the same $N$ and $M$ but different topology. On the other hand, it is also impossible to apply exhaustive attacks on a large-sized network, so a random attack is applied with many repeated runs instead.

For any node $i$, if its in- or out-degree does not satisfy Eq. (\ref{eq:ki}), edge rectification is needed. There are four possible edge rectification operations:
\begin{enumerate}
	\item If $k_{i}^{out}<\floor{M/N}$, then find another node $k$ with out-degree greater than $\ceil{M/N}$, and randomly pick one of its out-edges, $A_{k,l}$. Delete this edge $A_{k,l}$ and add an edge $A_{i,l}$. This increases $k_{i}^{out}$ by one and decreases $k_{k}^{out}$ by one.
	\item If $k_{i}^{out}>\ceil{M/N}$, then randomly pick one of its out-edges $A_{i,j}$, and find another node $k$ with out-degree less than $\floor{M/N}$. Delete this edge $A_{i,j}$ and add an edge $A_{k,j}$. This decreases $k_{i}^{out}$ by one and increases $k_{k}^{out}$ by one.
	\item If $k_{i}^{in}<\floor{M/N}$, then find another node $k$ with in-degree greater than $\ceil{M/N}$, and randomly pick one of its in-edges $A_{l,k}$. Delete this edge $A_{l,k}$ and add an edge $A_{l,i}$. This increases $k_{i}^{in}$ by one and decreases $k_{k}^{in}$ by one.
	\item If $k_{i}^{in}>\ceil{M/N}$, then randomly pick one of its in-edges $A_{j,i}$, and find another node $k$ with in-degree less than $\floor{M/N}$. Delete this edge $A_{j,i}$ and add an edge $A_{j,k}$. This decreases $k_{i}^{in}$ by one and increases $k_{k}^{in}$ by one.
\end{enumerate}

An execution of any of the above four rectifications is said to be an RER operation. Specifically, the RER strategy is defined as follows: pick a random operation of the four edge rectification operations, until the stop criterion is met. The stop criterion is either ``the maximum number of RER operations is reached'' or ``the network has satisfied the ENC''.

Given two networks, the one requiring less number of RER operations to exactly satisfy the ENC is said to be closer to satisfying ENC.

\section{Experimental Studies}
\label{sec:exp}

In this section, the RER strategy is applied to several different complex network topologies, including six synthetic and two real-world networks, to improve their robustness of controllability. The influences of RER is investigated on 1) modifying the networks' degree distributions towards satisfaction of ENC, 2) enhancing the robustness of controllability, and 3) enhancing the robustness of connectedness.

Six typical directed synthetic network models are adopted for simulation, namely the Erd{\"{o}}s--R{\'e}nyi random graph (ER) \cite{Erdos1964RG}, Newman--Watts small-world (SW) network \cite{Newman1999PLA}, generic scale-free (SF) network \cite{Pu2012PA,Goh2001PRL,Sorrentino2007CH}, \textit{q}-snapback network (QSN) \cite{Lou2018TCASI,Lou2019R}, random triangle network (RTN) \cite{Chen2019TCASII}, and random rectangle network (RRN) \cite{Chen2019TCASII}.

Two real-world networks are used for verification, namely the email-Eu-core (EE)\footnote{\url{https://snap.stanford.edu/data/email-Eu-core.html}} network and the Gnutella peer-to-peer (GP)\footnote{\url{https://snap.stanford.edu/data/p2p-Gnutella08.html}} network, which will be detailed in Section \ref{sub:rwn}.

In the following, the generation methods and parameters of the six synthetic networks are introduced.

\subsubsection{Erd{\"{o}}s--R{\'e}nyi Random Graph Networks} An ER network is generated as follows:
\begin{itemize}
	\item Start with $N$ isolated nodes.
	\item Pick up all possible pairs of nodes from the $N$ given nodes, denoted by $i$ and $j$ ($i\neq j$, $i,j=1,2,...,N$), once and once only. Connect each pair of nodes by a directed edge with probability $p_{RG}\in[0,1]$, where the edge has the same probability directing from $i$ to $j$, or $j$ to $i$.
\end{itemize}

Given the numbers of $N$ and $M$, let $p_{RG}=\frac{M}{N(N-1)}$. To exactly control the number of generated edges to be $M$, uniformly-randomly adding or removing edges can be performed. Here, when adding an edge, the direction can be random.

\subsubsection{Newman--Watts Small-world Networks} An SW network is generated as follows:
\begin{itemize}
	\item Start with a directed $N$-node loop having $K$ connected nearest-neighbors on each side of each node.
	\item Additional edges with random directions are added without removing any existing edges.
\end{itemize}

Set $K=2$ in the following, namely, a node $i$ is connected to its two nearest neighbors on each side, with nodes $i-1$, $i+1$, $i-2$ and $i+2$, via edges $A_{i-1,i}$, $A_{i,i+1}$, $A_{i-2,i}$ and $A_{i,i+2}$.

\subsubsection{Scale-Free Networks} An SF network is generated as follows:
\begin{itemize}
	\item Start with $N$ isolated nodes.
	\item A weight $w_{i}=(i+\theta)^{-\sigma}$ is assigned to node $i$, with $\sigma\in[0,1)$ and $\theta\ll N$.
	\item Two nodes $i$ and $j$ ($i\neq j$, $i,j=1,2,...,N$) are randomly picked from the pool with a probability proportional to the weights $w_i$ and $w_j$, respectively. Then, an edge $A_{ij}$ from $i$ to $j$ is added (if the two nodes are already connected, do nothing).
	\item Repeat Step 3), until $M$ edges have been added.
\end{itemize}

The resulting network has a power-law distribution $k^{-\gamma}$ with $\gamma=1+\frac{1}{\sigma}$, where $k$ is the degree variable, which is independent of $\theta$. Here, $\sigma$ is set to $0.999$, and thus $\gamma=2.001$.

\subsubsection{\textit{q}-Snapback Networks} Consider a \textit{q}-snapback network (QSN) with only one layer $r_{QSN}$ for simplicity. This QSN is generated as follows:
\begin{itemize}
	\item Start with a directed chain of $N$ nodes, where each node $i$ ($i=1,2,...,N-1$) has an edge $A_{i,i+1}$.
	\item For each node $i=r_{QSN}+1,\, r_{QSN}+2, \ldots, N$, it connects backward to the previously-appeared nodes $i-l\times r_{QSN}$ ($l=1, 2, \ldots, \floor{i/r_{QSN}}$), with the same probability $q\in[0,1]$.
\end{itemize}

In the following experimental study, $r_{QSN}$ is set to $2$. Given $N=1000$ and $M=5000$, $q$ is estimated to be $0.008$ for fair comparisons. To exactly generate $M$ edges, uniformly-randomly edge-adding with random direction should be applied.

\subsubsection{Random Triangle Networks} Triangular structure, which has been observed benefit to the robustness of controllability \cite{Lou2018TCASI} and network stability \cite{Schultz2014NJP, Nitzbon2017NJP}, is frequently observed in real-life situations.

A directed random triangle network (RTN) is generated as follows:
\begin{itemize}
	\item Start with $N-3$ isolated nodes, with the other 3 nodes connected in a directed triangle.
	\item Randomly pick up two nodes, $i$ and $j$, without edge $A_{ij}$ or $A_{ji}$ (otherwise, do nothing). Then, randomly pick up a node $k$ from all the neighbors of node $j$. If there is an edge $A_{jk}$, then add two edges $A_{ij}$ and $A_{ki}$; otherwise (e.g., with an edge $A_{kj}$), add two edges $A_{ji}$ and $A_{ik}$.
	\item Repeat Step 2), until $M$ edges have been added.
\end{itemize}

\subsubsection{Random Rectangle Networks} The above directed RTN is extended to a random rectangle network (RRT), as follows:
\begin{itemize}
	\item Start with $N-4$ isolated nodes, and the other 4 nodes are connected in a directed rectangle.
	\item Randomly pick up three nodes, $i$, $j$ and $k$, without edges between any pair of them (otherwise, do nothing). Then, randomly pick up a node $w$ from the neighbors of node $k$. If there is an edge $A_{kw}$, then add edges $A_{wi}$, $A_{ij}$, and $A_{jk}$; otherwise (e.g., with an edge $A_{wk}$), add edges $A_{ki}$, $A_{ij}$, and $A_{jw}$.
	\item Repeat Step 2), until $M$ edges have been added.
\end{itemize}

Since at each time step, two edges are added into RTN, and three edged are added into RRN, uniformly-randomly adding or removing edges can be performed to control the number of edges exactly.

In the simulation below, the network size is $N=1000$ with average out-degree $\langle K^{out}\rangle=5$, i.e., $M=5000$. To minimize the influence of stochasticity, for each configuration with the given $N$, $M$ and the number RER operations, repeat $30$ independent runs (yielding $30$ instances). For each instance, random attack is performed $50$ times independently. Thus, each statistic datum plotted in Figs. \ref{fig:nd(sc)_vs_pn}, \ref{fig:ho_vs_pn}, \ref{fig:hi_vs_pn} and \ref{fig:t_vs_network} is obtained by averaging $1500$ runs. Simulation results with different network sizes of $N=\{500,2000\}$, and different average out-degrees of $\langle K^{out}\rangle=\{3,8\}$, are shown in Figs. S6--S13 of the SI.

\subsection{Towards Satisfaction of the ENC}
\label{sub:t_enc}

\begin{figure}[htbp]
	\centering
	\includegraphics[width=.45\linewidth]{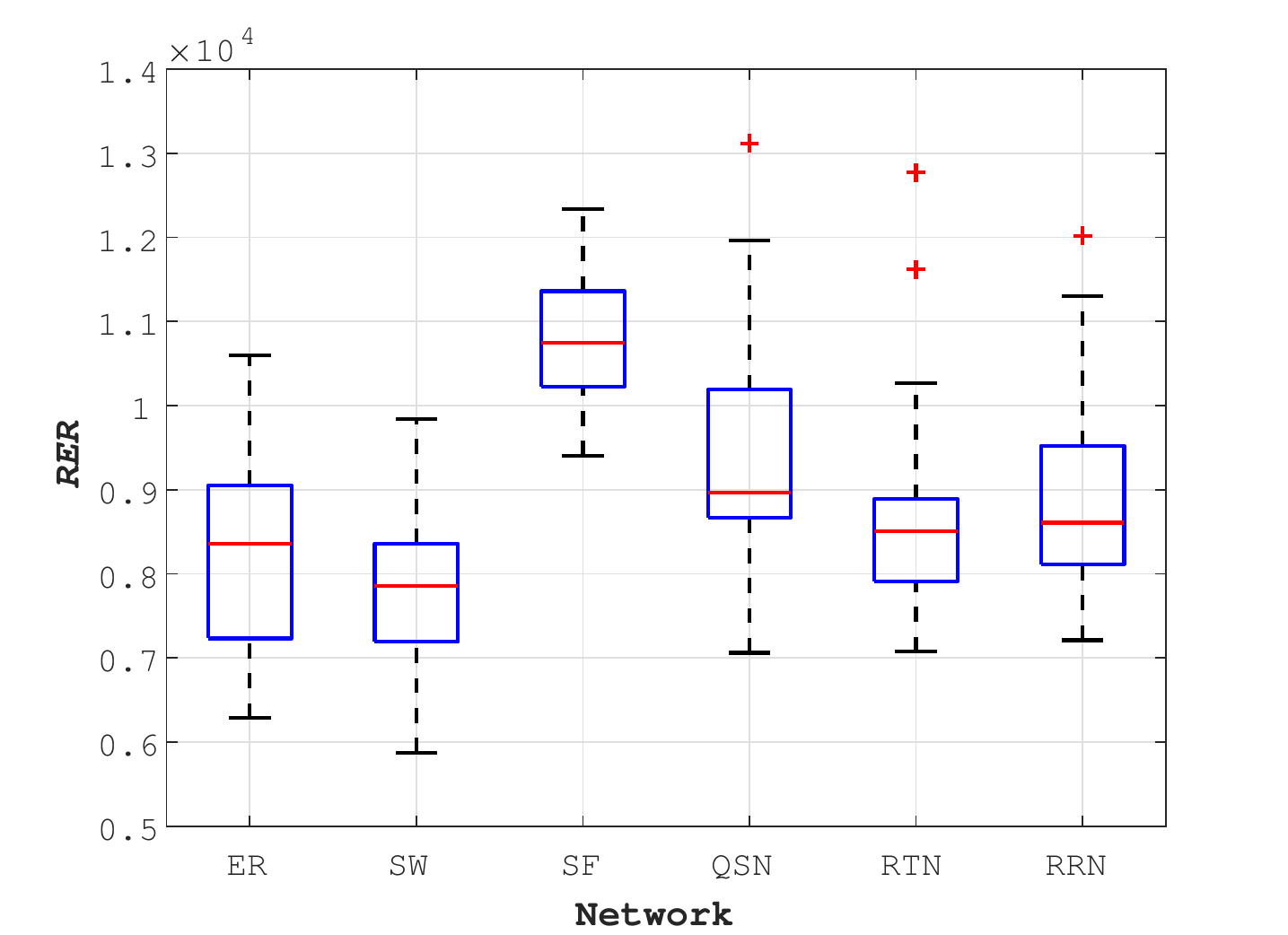}
	\caption{[Color online] Number of RER operations to rectify a network to satisfy the ENC. The network configuration has $N=1000$ and $M=5000$; the number of repeated runs is $100$. In a boxplot, the blue box indicates that the central $50\%$ samples lie within this section; the red bar inside the box is the median; the upper and lower black bars are the greatest and least values, excluding outliers; and the red pluses represent the outliers. }
	\label{fig:rer_vs_network}
\end{figure}

Figure \ref{fig:rer_vs_network} presents the boxplot of the needed number of RER operations for a network to exactly satisfy the ENC. It can be seen from the figure that, for a network configuration with $N=1000$ and $M=5000$, ER, SW, QSN, RTN, and RRN require a median of rounds of $0.8\times10^4$ to $0.9\times10^4$ for rectification, while SF requires rounds of $1\times10^4$ to $1.1\times10^4$. SW is the closest to satisfying the ENC, while SF is the farthest. Referring to Fig. \ref{fig:nd(sc)_vs_pn}(A), SW shows the best robustness of controllability, while SF shows the worst. Or, simply put, a network with better robustness of controllability needs less number of RER operations towards satisfaction of the ENC. Given different values of $N$ and $M$, the needed number of RER operations for ER and SF can be found in Fig. S14 of the SI.

\subsection{Towards Optimal Robustness of Controllability}
\label{sub:t_orc}

Now, the change of robustness of controllability is studied, as the number of RER operations changes.

First, note from Fig. \ref{fig:rer_vs_network} that it requires about ten thousands of RER operations for a network with $N=1000$ and $M=5000$ to satisfy the ENC.

Then, to see the changes, the following four situations are compared: 1) no RER operation is implemented, i.e., the networks are generated via their generation methods without any modification; 2) $1000$ RER operations are implemented, which are around one tenth of the number of the needed RER operations to exactly satisfy the ENC. The RER operations are repeatedly implemented on the original networks for $1000$ times, and the controllability robustness of the resultant networks is then investigated. 3) $5000$ RER operations are implemented, which are around one half of the number of operations to exactly satisfy the ENC; 4) unlimited RER operations (denoted by Inf) until the ENC is exactly satisfied. Note that the network structure can be significantly changed, e.g., given an SF with $1000$ RER, the resultant network is no longer scale-free (see Fig. \ref{fig:er_sf_pKo}(B)), but it is still referred to as an SF network for its original topology is scale-free.

\begin{figure*}[htbp]
	\centering
	\includegraphics[width=\linewidth]{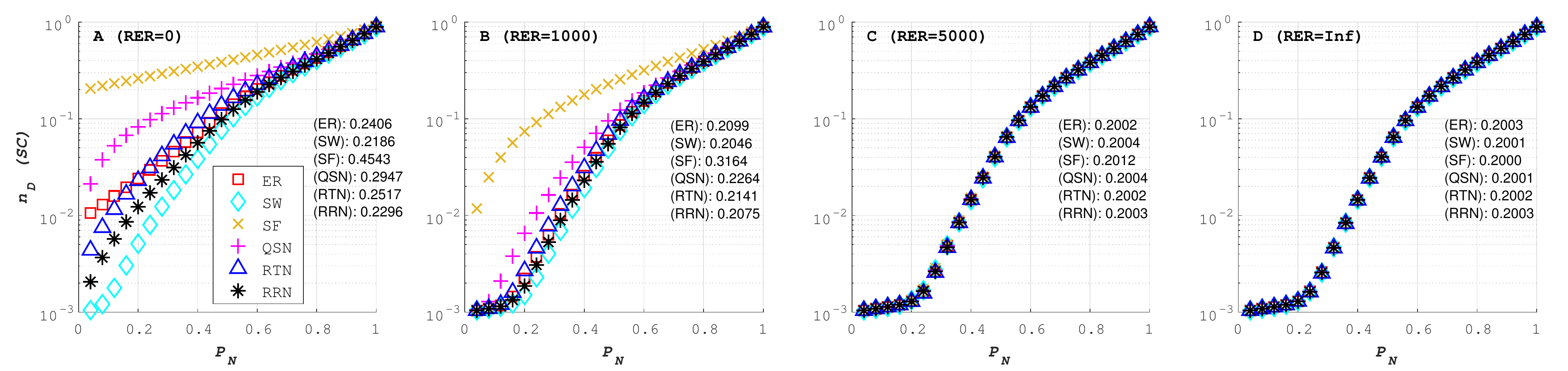}
	\caption{[Color online] Robustness of \textit{structural} controllability of the six networks: (A) without any rectification; (B) with $1000$ RER operations; (c) with $5000$ RER operations; (D) with RER operations until ENC is exactly satisfied. $n_D$ represents the density of controlled-nodes calculated by Eq. (\ref{eq:nd}); $P_N$ represents the proportion of removed nodes. Each curve is obtained by averaging $1500$ independent runs. The $\langle R_c\rangle$ values are placed beside the curves. (Robustness of \textit{exact} controllability can be found in Fig. S4 of the SI.)}
	\label{fig:nd(sc)_vs_pn}
\end{figure*}

Figure \ref{fig:nd(sc)_vs_pn} shows that the robustness of structural controllability of the six networks is enhanced as the number of RER operations increases. Figure \ref{fig:nd(sc)_vs_pn}(A) shows different curves of controllability; Fig. \ref{fig:nd(sc)_vs_pn}(B) shows that the robustness of all networks is improved and the difference of their performances becomes smaller; in Figs. \ref{fig:nd(sc)_vs_pn}(C) and (D), the robustness is significantly enhanced and the differences among the curves become indistinguishable, for which although there is no guarantee by theory that the rectified networks have optimal robustness of controllability, it is obvious that the robustness is significantly enhanced.

\begin{figure*}[htbp]
	\centering
	\includegraphics[width=\linewidth]{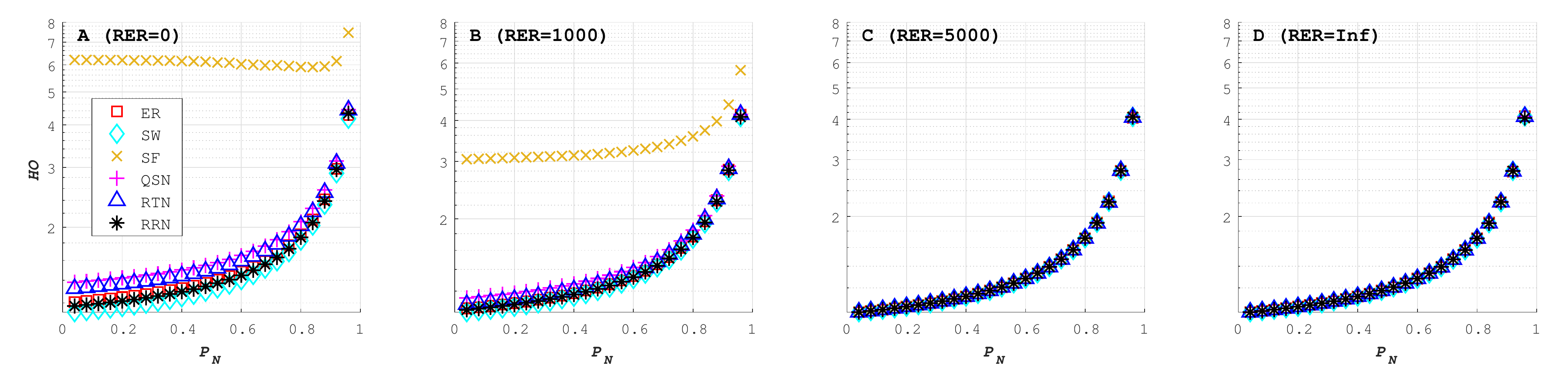}
	\caption{[Color online] Heterogeneity of out-degree (HO) against the proportion of removed nodes ($P_N$) of the six networks: (A) without any rectification; (B) with $1000$ operations; (c) with $5000$ RER operations; (D) with RER operations until ENC is exactly satisfied. Each curve is obtained by averaging $1500$ independent runs.}
	\label{fig:ho_vs_pn}
\end{figure*}

\begin{figure*}[htbp]
	\centering
	\includegraphics[width=\linewidth]{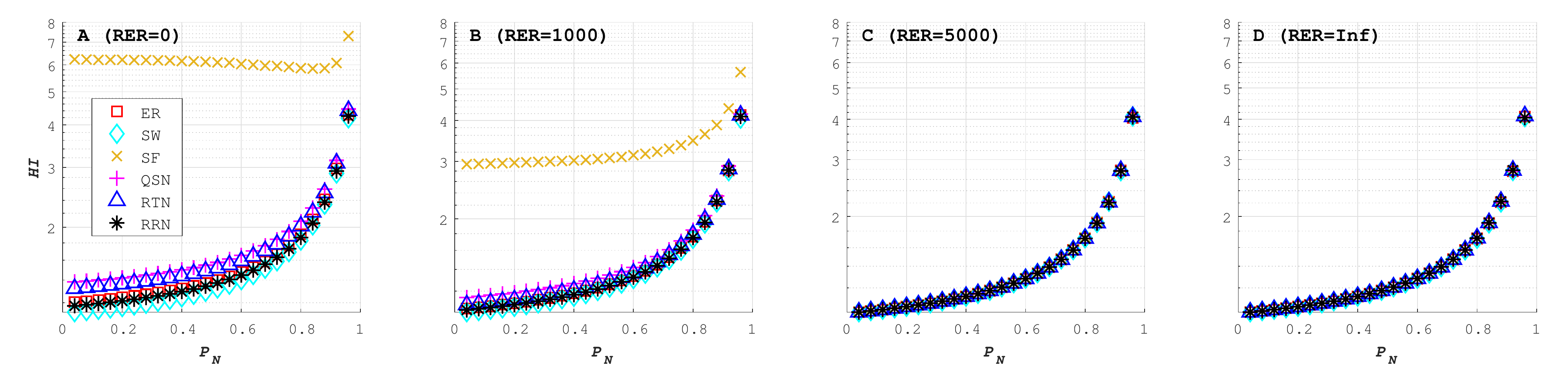}
	\caption{[Color online] Heterogeneity of in-degree (HI) against the proportion of removed nodes ($P_N$) of the six networks: (A) without any rectification, (B) with $1000$ operators; (c) with $5000$ RER operations; (D) with RER operations until ENC is exactly satisfied. Each curve is obtained by averaging $1500$ independent runs.}
	\label{fig:hi_vs_pn}
\end{figure*}

Figures \ref{fig:ho_vs_pn} and \ref{fig:hi_vs_pn} show the change of heterogeneity of out- and in-degrees (HO and HI), against the proportion of removed nodes. Here, network node-degree heterogeneity is considered, which is characterized by $\text{H}=\langle k^2\rangle/\langle k\rangle ^2$, where $\langle\cdot\rangle$ denotes the average. When $k$ is in-degree, it returns to HI; when $k$ is out-degree, it returns to HO. It can be seen from the figures that both HO and HI increase as nodes are gradually removed. As can also be seen from Figs. \ref{fig:ho_vs_pn}(A) and \ref{fig:hi_vs_pn}(A), the original SF network has the highest HO and HI, suggesting that low HO and HI values imply better robustness of controllability. As the number of RER operations increases (Figs. \ref{fig:ho_vs_pn}(B,C) and \ref{fig:hi_vs_pn}(B,C)), both HO and HI are reduced, finally resulting in extremely-homogeneous networks (Figs. \ref{fig:ho_vs_pn}(D) and \ref{fig:hi_vs_pn}(D)), which show the best robustness of controllability.

Since connectedness is also important in the regard of network controllability and other issues, the proportion of node-removals needed to disconnect a network, namely, the minimum proportion of nodes to be removed in order to break the network into disjoint components, under \textit{random} attacks, is examined next.

\begin{figure*}[htbp]
	\centering
	\includegraphics[width=\linewidth]{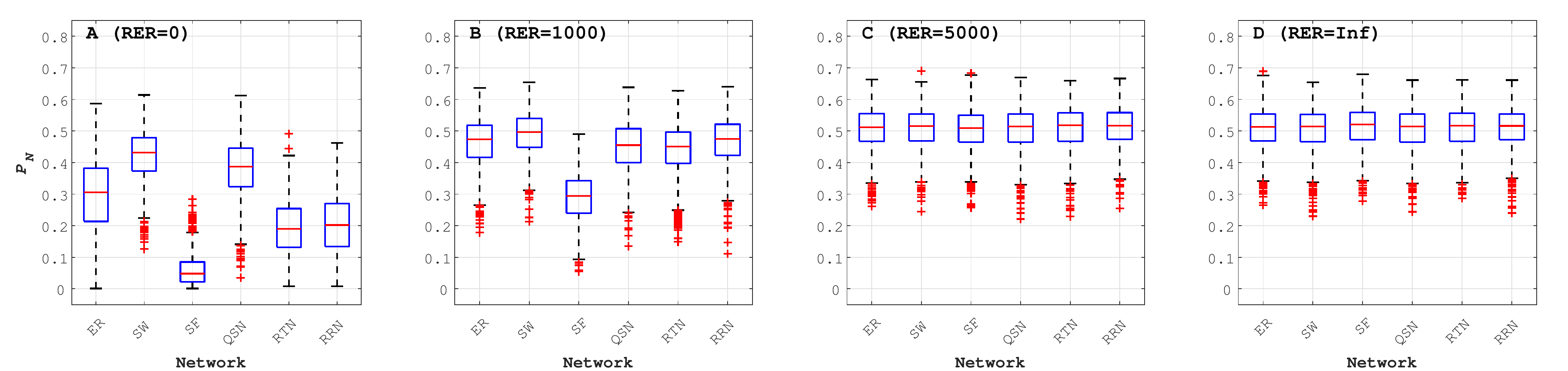}
	\caption{[Color online] Proportion of random node-removals (denoted by $P_N$) to disconnect a network: (A) without any rectification; (B) with $1000$ operations; (c) with  $5000$ RER operations; (D) with RER operations until ENC is exactly satisfied.}
	\label{fig:t_vs_network}
\end{figure*}

As can be seen from Fig. \ref{fig:t_vs_network}(A), different networks show different behaviors against random attacks, where SF is the easiest to be disconnected among the six networks, while SW is the hardest. This phenomenon is consistent with what was presented in Fig. \ref{fig:rer_vs_network}(A), where SF shows the worst robustness against attacks, while SW has the best. In Fig. \ref{fig:t_vs_network}(B), the proportion of node-removals to disconnect all networks is increased, meaning that $1000$ RER operations improve the connectedness of all networks against attacks. Finally, in Fig. \ref{fig:t_vs_network}(C,D), the value of $P_N$ is further increased. Fig. \ref{fig:t_vs_network} demonstrates that the robustness of network connectedness is improved as the number of RER operations increases.

\subsection{Extensive Simulations on ER and SF}
\label{sub:er_n_sf}

\begin{figure}[htbp]
	\centering
	\includegraphics[width=.75\linewidth]{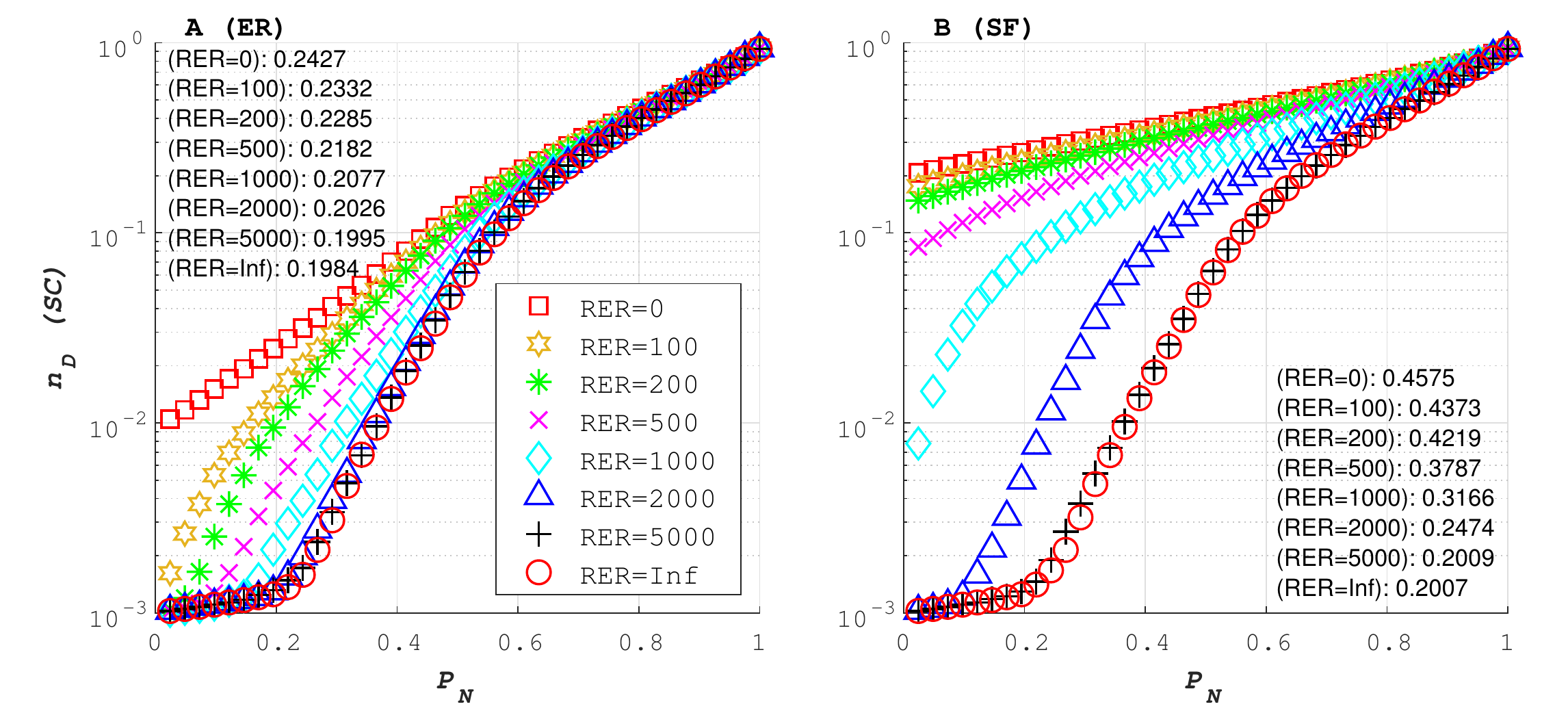}
	\caption{[Color online] Robustness of \textit{structural} controllability of (A) ER, and (B) SF. $n_D$ represents the density of controlled-nodes calculated by Eq. (\ref{eq:nd}); $P_N$ represents the proportion of removed nodes. The detailed $\langle R_c\rangle$ values are attached beside the curves. Each curve is obtained by averaging $100$ independent runs. (Robustness of \textit{exact} controllability can be found in Fig. S5 of the SI.)}
	\label{fig:er_sf_ndSC}
\end{figure}

Next, only ER and SF are discussed. A small step length of RER increase is set, such that the subtle influences of RER on the robustness of both controllability and connectedness can be clearly evaluated.

Figure \ref{fig:er_sf_ndSC} shows the curves of the robustness of controllability of ER and SF, as the number of RER operations increases, namely, with RER $=\{0,100,200,500,1000,2000,5000,\text{Inf}\}$. The results are obtained by averaging $100$ repeated runs. It is obvious that increasing the number of RER operations significantly enhances the robustness of controllability. The curves are distinguishable even when $100$ RER operations are performed on the networks, showing that the operations have clear impacts.

\begin{figure}[htbp]
	\centering
	\includegraphics[width=.75\linewidth]{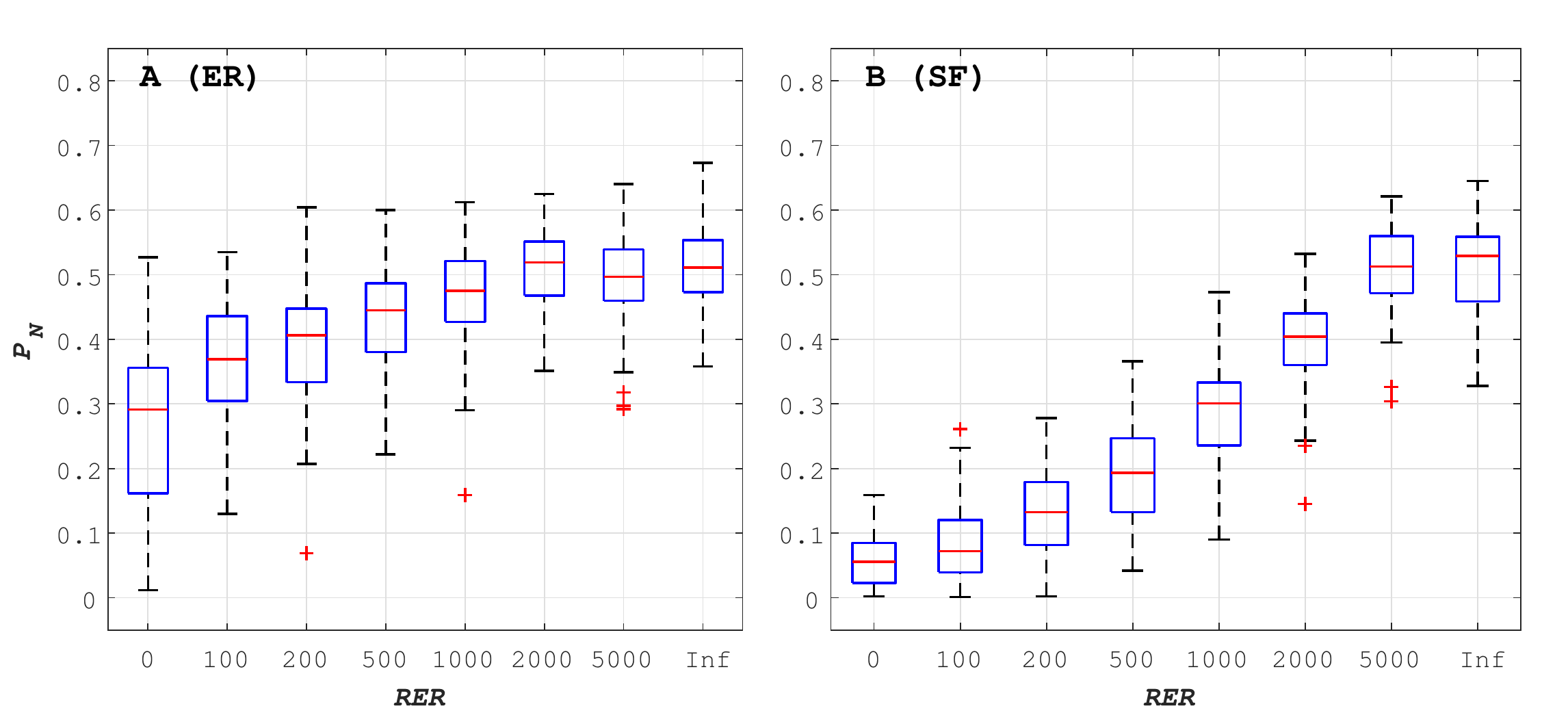}
	\caption{[Color online] Proportion of random node-removals (denoted by $P_N$) against the number of RER operations to disconnect a network: (A) ER, and (B) SF.}
	\label{fig:er_sf_t_vs_r}
\end{figure}

Figure \ref{fig:er_sf_t_vs_r} shows the proportion of random node-removals to disconnect a network, against the number of RER operations. The data are the averaged results of $100$ repeated runs. In this figure, a higher $P_N$ value means more node-removals is needed to disconnect the network, namely, the network has better robustness of connectedness. For both ER and SF, the robustness of connectedness is improved as the number RER operations increases.

\begin{figure}[htbp]
	\centering
	\includegraphics[width=.75\linewidth]{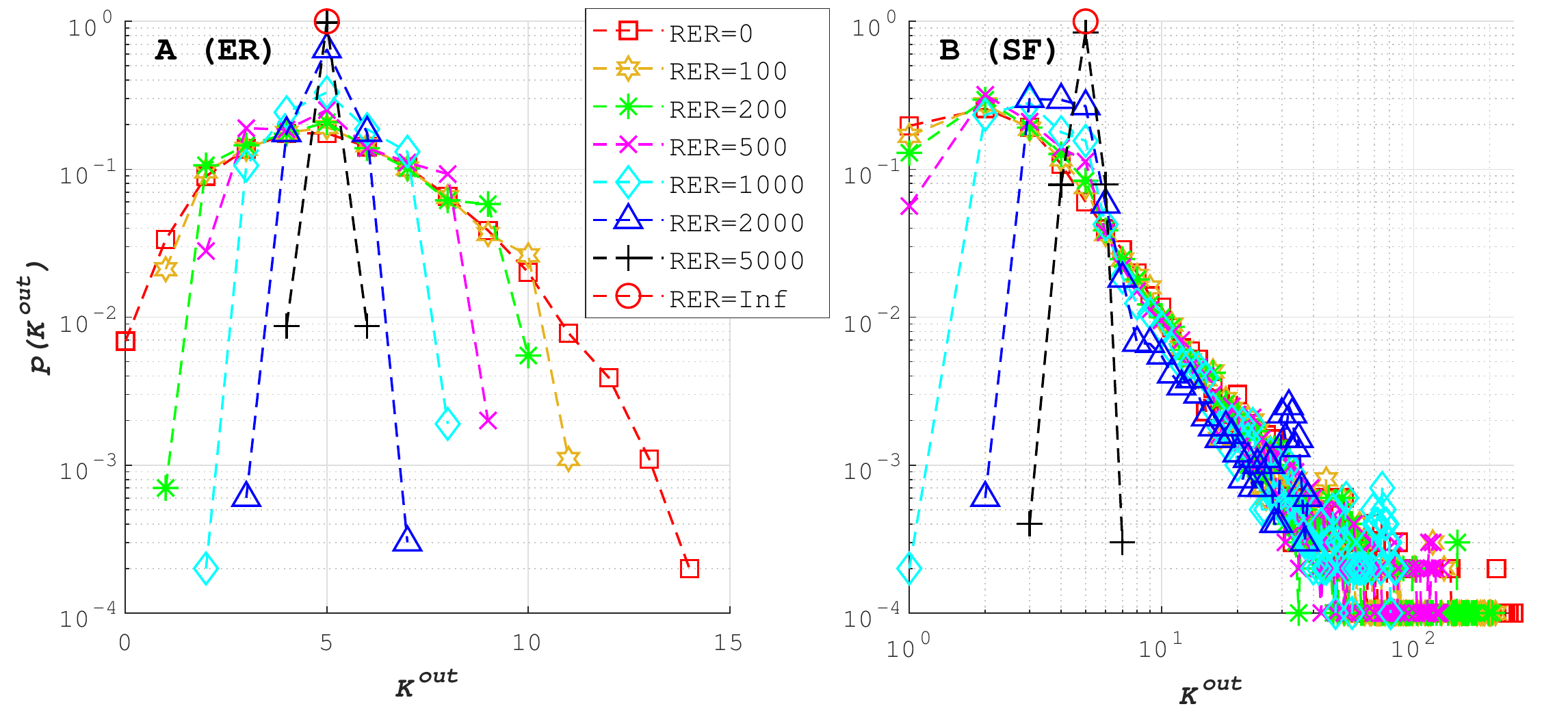}
	\caption{[Color online] Out-degree distribution changes as the number of RER operations increases: (A) ER, and (B) SF. (In-degree distribution can be found in Fig. S15 of the SI.)}
	\label{fig:er_sf_pKo}
\end{figure}

As shown in Figs. \ref{fig:ho_vs_pn} and \ref{fig:hi_vs_pn}, the operation of RER makes a network gradually become homogeneous. Fig. \ref{fig:er_sf_pKo} shows the change of out-degree distributions of ER and SF. Fig. \ref{fig:er_sf_pKo}(A) shows that the out-degree distribution of the original ER is Poisson, but as the number of RER operations increases, the out-degree distribution becomes concentrated at $M/N$ ($M/N=5$ in the figure). Fig. \ref{fig:er_sf_pKo}(B) shows that the original power-law distribution of SF also concentrates at $M/N$. Unlimited number of RER operations make both ER and SF (and any other network) become extremely homogeneous.

\subsection{Two real-world Networks}
\label{sub:rwn}

\begin{table}[htbp]
	\centering
	\caption{Parameters and descriptions of the two real-world networks (the number of edges $M$ of EE is $25571$ in \cite{SNAP2014}; after discarding self-loops, it becomes $24929$)}
	\begin{tabular}{|l|l|c|c|}
		\hline
		\multicolumn{1}{|c|}{Network} & \multicolumn{1}{c|}{Description} & $N$ & $M$ \\ \hline
		\begin{tabular}[c]{@{}l@{}}email-Eu-core\\ (EE)\end{tabular} & \begin{tabular}[c]{@{}l@{}}a directed network generated \\ using email data from \\ European research institution\end{tabular} & $1005$ & $24929$ \\ \hline  
		\begin{tabular}[c]{@{}l@{}}Gnutella\\ peer-to-peer\\ (GP)\end{tabular} & \begin{tabular}[c]{@{}l@{}}a directed network generated\\ by snapshots of Gnutella \\ peer-to-peer file sharing \\ network in August 2002\end{tabular} & $6301$ & $20777$ \\ \hline
	\end{tabular}\label{tab:rwn}
\end{table}

\begin{figure}[htbp]
	\centering
	\includegraphics[width=.75\linewidth]{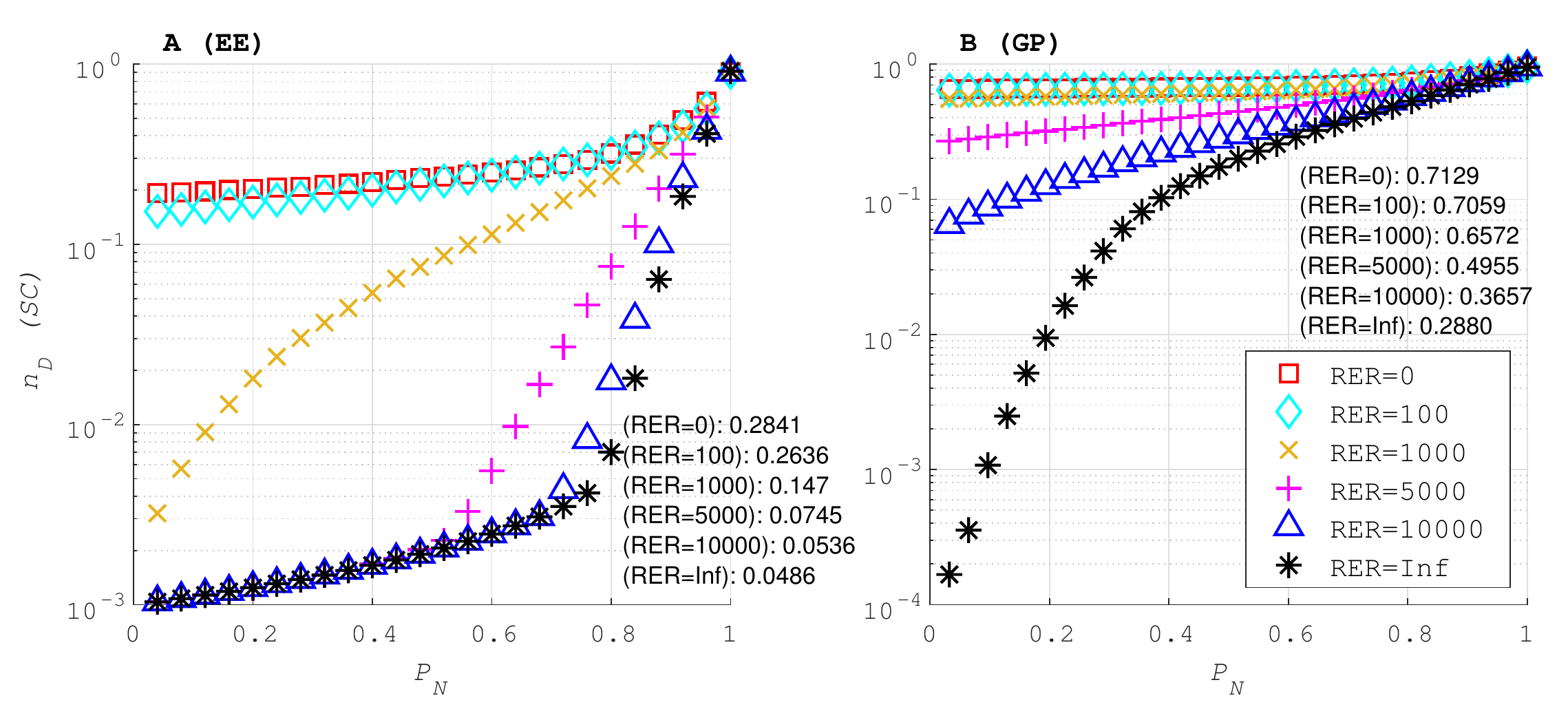}
	\caption{[Color online] Robustness of \textit{structural} controllability of (A) EE, and (B) GP. $n_D$ represents density of controlled-nodes calculated by Eq. (\ref{eq:nd}); $P_N$ represents the proportion of removed nodes. The detailed $\langle R_c\rangle$ values are attached beside the curves. Each curve is obtained by averaging $100$ independent runs.}
	\label{fig:rwn_ndsc_vs_pn}
\end{figure}

\begin{figure}[htbp]
	\centering
	\includegraphics[width=.75\linewidth]{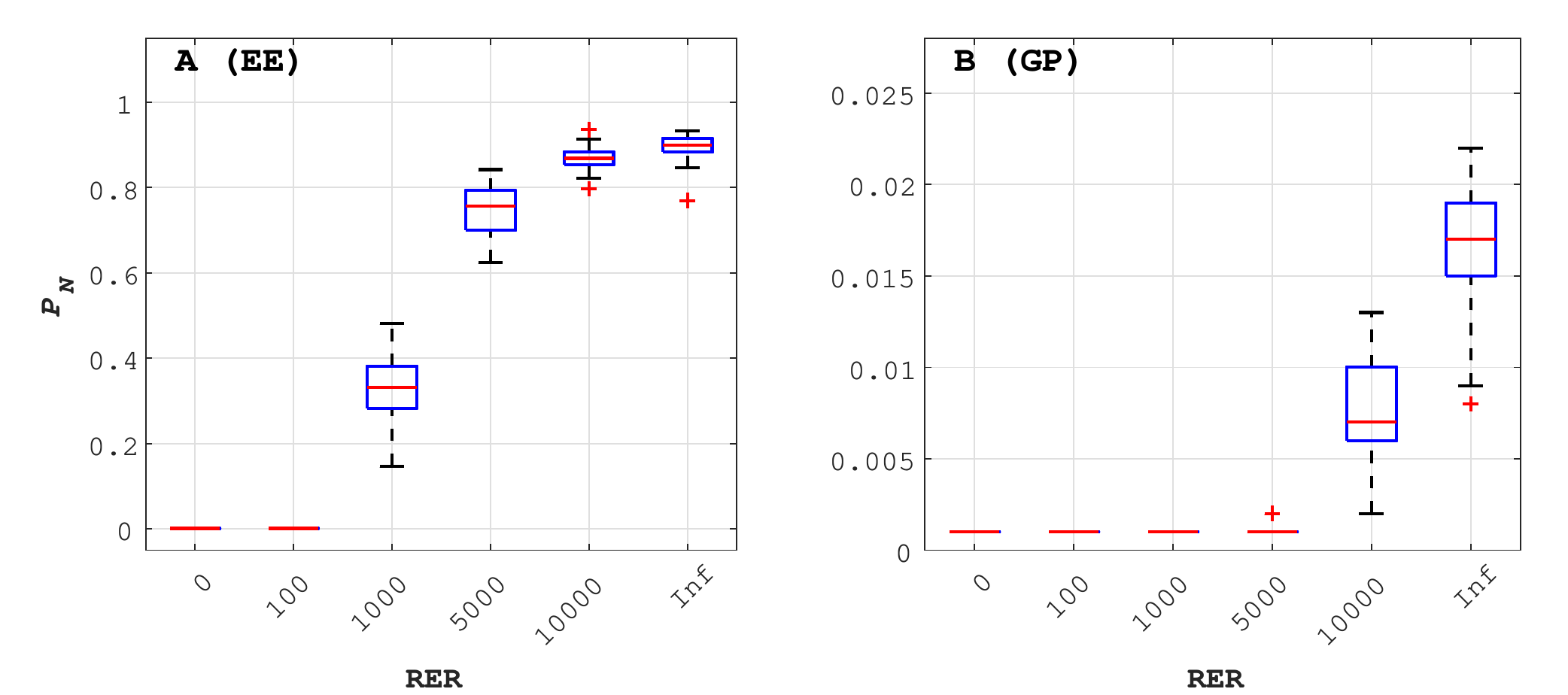}
	\caption{[Color online] Proportion of random node-removals (denoted by $P_N$) against the number of RER operations to disconnect a network: (A) EE, and (B) GP.}
	\label{fig:rwn_pn_vs_rer}
\end{figure}

Consider two real-world networks, namely email-Eu-core (EE) network and Gnutella peer-to-peer (GP) network \cite{SNAP2014}. Their parameters and brief descriptions are presented in Table \ref{tab:rwn}.

The original EE and GP (with RER$=0$) are compared to the settings with RER $=\{100,1000,5000,10000,\text{Inf}\}$. The controllability curves are shown in Fig. \ref{fig:rwn_ndsc_vs_pn}. A phenomenon similar to the simulations on synthetic networks is observed: the robustness of controllability is improved as the number of RER operations increases. The controllability curves of the original EE and GP (with RER$=0$) are far away from the curves that exactly satisfy the ENC (with RER$=\text{Inf}$). This clearly shows that real-world networks are far away from having optimal robustness of controllability.

In Fig. \ref{fig:rwn_pn_vs_rer}, the boxplot presents the proportion of node-removals needed to break the network into disjoint components, for EE and GP respectively. It can be observed that the original real-world networks are quite fragile, but after rectified with RER operations, their robustness of connectedness is significantly improved.

\subsection{Discussions}
\label{sub:disc}

\begin{table*}[htbp]
	\centering
	\caption{Changes of basic network features, when RER is set to $0$, $1000$, and Inf, respectively. The compared features include average degree ($\langle k\rangle$), average path length (\textit{apl}), average (node) betweenness centrality (\textit{abc}), clustering coefficient (\textit{cc}), number of basic loops (\textit{nbl}), heterogeneity of out-degrees (\textit{ho}), and heterogeneity of in-degrees (\textit{hi}).}
	\begin{tabular}{|c|c|c|c|c|c|c|c|}
		\hline
		Net&\multicolumn{3}{c|}{ER ($N=1000$)}&\multicolumn{3}{c|}{SF ($N=1000$)}\\ \hline
		RER&$0$& $1000$&Inf&$0$& $1000$&Inf \\ \hline
		$\langle k\rangle$&5&5&5&5&5&5\\ \hline
		\textit{apl}&Inf&4.3365&4.2943&Inf&3.8097&4.2943\\ \hline
		\textit{abc}&3.4E+03&3.3E+03&3.2E+03&2.4E+03&2.8E+03&3.2E+03 \\ \hline
		\textit{cc}&0.0051&0.0044&0.0041&0.0947&0.0334&0.0040 \\ \hline
		\textit{nbl}&2039.0&2170.9&2201.2&1576.3&1832.9&2187.3 \\ \hline
		\textit{ho}&1.1992&1.0532&1 &6.2678&3.0727&1 \\ \hline
		\textit{hi}&1.1984&1.0533&1 &6.1320&2.9580&1 \\ \hline\hline
		
		Net&\multicolumn{3}{c|}{EE ($N=1005$)}&\multicolumn{3}{c|}{GP ($N=6301$)}\\ \hline
		RER&$0$& $1000$&Inf&$0$& $1000$&Inf \\ \hline
		$\langle k\rangle$&5&5&5&5&5&5 \\ \hline
		\textit{apl}&Inf&4.3365&4.2943&Inf&3.8097&4.2943 \\ \hline
		\textit{abc}&3.4E+03&3.3E+03&3.2E+03&2.4E+03&2.8E+03&3.2E+03 \\ \hline
		\textit{cc}&0.0051&0.0044&0.0041&0.0947&0.0334&0.0040 \\ \hline
		\textit{nbl}&2039.0&2170.9&2201.2&1576.3&1832.9&2187.3 \\ \hline
		\textit{ho}&1.1992&1.0532&1 &6.2678&3.0727&1 \\ \hline
		\textit{hi}&1.1984&1.0533&1 &6.1320&2.9580&1 \\ \hline
	\end{tabular}  \label{tab:cmp}
\end{table*}

It can be observed from Fig. \ref{fig:er_sf_ndSC} that, when unlimited RERs are implemented, the resultant extremely-homogeneous ER and SF networks are very similar in terms of controllability robustness. However, Fig. \ref{fig:rwn_ndsc_vs_pn} shows that, when unlimited RERs are implemented to EE and GP, their resultant networks have very different controllability robustness.

Table \ref{tab:cmp} summarizes the changes of some basic features, where the number of RER is set to $0$, $1000$, and Inf, respectively. The compared features include average degree, average path length, average (node) betweenness centrality, clustering coefficient, number of basic loops, heterogeneity of out-degrees, and heterogeneity of in-degrees.

Obviously, RER does not change the average degree of a network, but shortens the average path length of all networks from Inf (when there is at least one node $i$ that is unreachable by any other node $j$) to a small value. RER slightly decreases the average betweenness centrality of ER, but increases that of SF, EE, and GP. RER decreases the clustering coefficients, but increases the number of basic loops, for all the four networks. Finally, as expected, RER decreases both in- and out-degree heterogeneities of all networks.

\begin{table*}[htbp]
	\centering
	\caption{Basic network features, when RER is set to Inf. The compared features include average degree ($\langle k\rangle$), average path length (\textit{apl}), average (node) betweenness centrality (\textit{abc}), clustering coefficient (\textit{cc}), number of basic loops (\textit{nbl}), heterogeneity of out-degrees (\textit{ho}), and heterogeneity of in-degrees (\textit{hi}).}
	\begin{tabular}{|c|c|c|c|c|c|c|}
		\hline
		Net & ER ($N=1005$) & SF ($N=1005$) & EE ($N=1005$) & ER ($N=6301$)& SF ($N=6301$) & GP ($N=6301$) \\ \hline
		RER                & Inf      & Inf      & Inf      & Inf      & Inf      & Inf      \\ \hline
		$\langle k\rangle$ & 24.80    & 24.80    & 24.80    & 3.30     & 3.30     & 3.30     \\ \hline
		apl                & 2.4954   & 2.4957   & 2.5132   & 7.0693   & 7.0677   & 7.0686   \\ \hline
		abt                & 1.50E+03 & 1.50E+03 & 1.52E+03 & 3.82E+04 & 3.82E+04 & 3.82E+04 \\ \hline
		cc                 & 0.0238   & 0.0238   & 0.0410   & 0.0003   & 0.0004   & 0.0004   \\ \hline
		nlp                & 12388.7  & 12242.7  & 12344.0  & 7716.0   & 7813.0   & 7754.0   \\ \hline
		ho                 & 1.0003   & 1.0003   & 1.0003   & 1.0192   & 1.0192   & 1.0192   \\ \hline
		hi                 & 1.0003   & 1.0003   & 1.0003   & 1.0192   & 1.0192   & 1.0192   \\ \hline
	\end{tabular}   \label{tab:cmp2}
\end{table*}

It is notable that unlimited RERs make the basic features of ER and SF very close to each other. This phenomenon verifies that RER can rectify any networks towards a same structure, or very similar ones, no matter what their original topologies are. However, this is not observed from EE and GP. Basically, due to the different sizes and average degrees of EE and GP, the basic features of these networks cannot be rectified uniformly like that of ER and SF with same $N$ and $M$ values. For example, the numbers of basic loops are very different in EE and GP, when ERE is Inf. As already observed, the multi-loop structure is very beneficial to controllability robustness \cite{Lou2018TCASI,Lou2019R,Chen2019TCASII}.

Table \ref{tab:cmp2} further verifies that, no matter what the original topology is, RER rectifies networks towards the same extremely-homogeneous structure, or very similar ones. Here, both ER and SF are set with $N=1005$ and $\langle k\rangle=24.80$, the same configuration as that of EE. After unlimited RER operations, most basic features of the three networks become very similar to each other, except a small difference in clustering coefficients. When ER and SF are set to the same configuration as GP ($N=6301$ and $\langle k\rangle=3.30$), all the basic features of the three networks become very similar.

\section{Conclusions}
\label{sec:end}

This paper presents a search for the network configuration with optimal robustness of controllability against random node-removal attacks. Since analytical approach seems impossible at least in the present time, the exhaustive attack strategy that employs all possible attack sequences is applied. Since this too is an intractable attempt even for the numerical approach, the work is performed on some very small-size networks. This nevertheless yields clear determined patterns of optimal solutions, which leads to an efective empirical necessary condition (ENC), indicating that the optimal instance of network configuration should be extremely homogeneous. This important finding of the present paper is consistent with some earlier observations reported in the literature. ENC rules out the network instances that would not be candidates having optimal robustness of controllability. A random edge rectification (RER) strategy is then proposed to rectify synthetic and real-world networks towards exact satisfaction of the ENC, which also provides a way to enhance the robustness of controllability. The observed ENC may be useful in designing future network models. The phenomenon observed in this paper has an important implication that real-world networks as well as the commonly-used synthetic models are actually far away from the topologies with optimal robustness of controllability. Future work along the same line may be extended to other scenarios of malicious attacks, e.g., edge-removal and intentional attacks.






\def\bibindent{1em}

\includepdf[page=-]{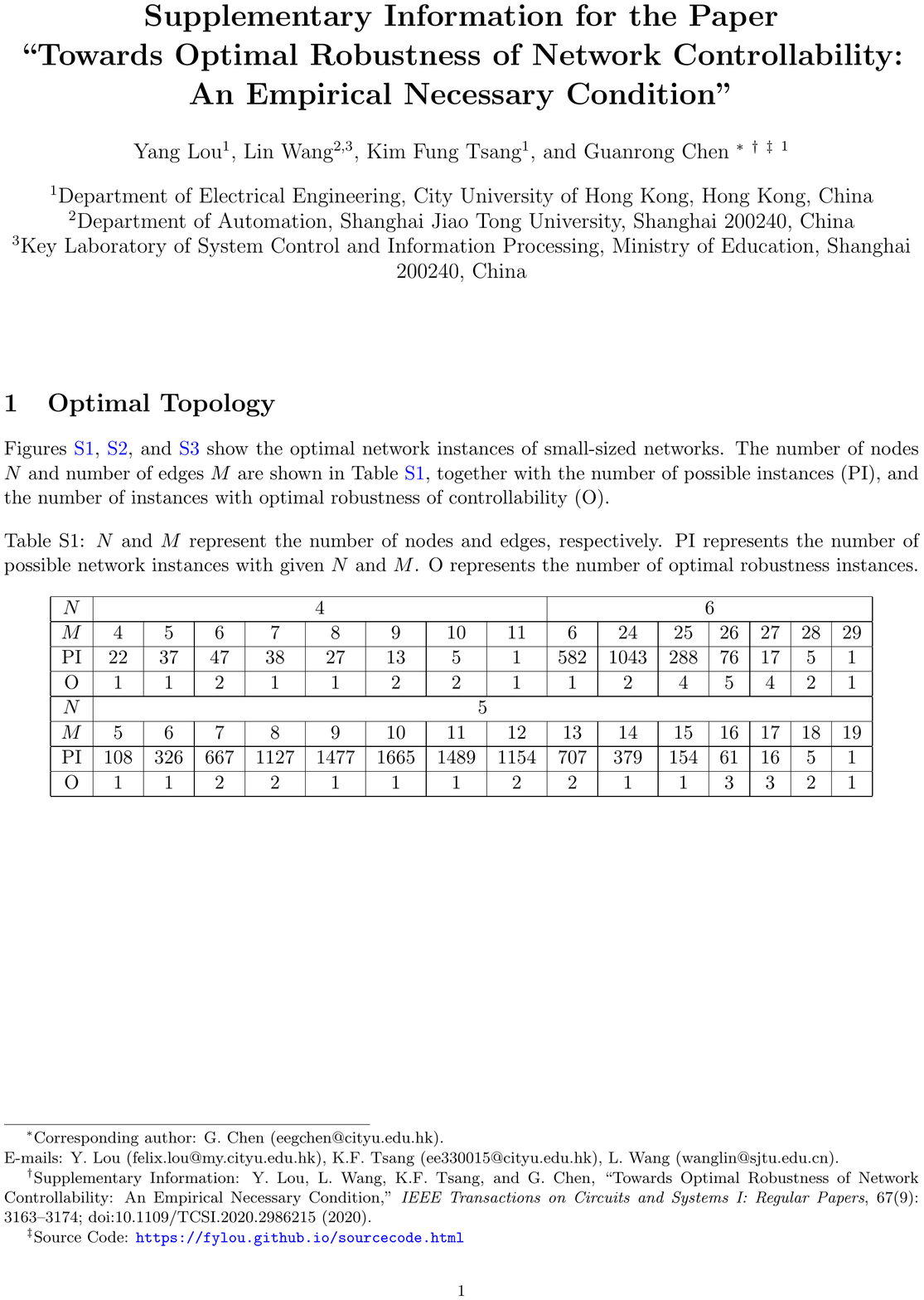}

\end{document}